\documentclass[runningheads]{llncs}
\usepackage[normalem]{ulem}
\usepackage[title]{appendix}
\usepackage{amssymb}
\usepackage{makecell}
\usepackage{graphicx}
\begin{document}
\title{Comparing Psychometric and Behavioral Predictors of Compliance During Human-AI Interactions
\thanks{Part of the effort behind this work was sponsored by the Defense Advanced Research Projects Agency (DARPA) under contract number W911NF2010011. The content of the information does not necessarily reflect the position or the policy of the U.S. Government or the Defense Advanced Research Projects Agency, and no official endorsements should be inferred. }
}
\titlerunning{Psychometric Vs. Behavioral Predictors of Compliance}
%
\author{ 
Nikolos Gurney\inst{1}\orcidID{0000-0003-3479-2037} \and 
David V. Pynadath\inst{1,2}\orcidID{0000-0003-2452-4733} \and
Ning Wang\inst{1,2}
}
\authorrunning{N. Gurney et al.}
%
\institute{ Institute for Creative Technologies \\
\email{gurney,pynadath,nwang@ict.usc.edu} \and
   Viterbi School of Engineering, Computer Science Department\\
   University of Southern California, Los Angeles, CA
  }
\maketitle              
\begin{abstract}
Optimization of human-AI teams hinges on the AI's ability to tailor its interaction to individual human teammates. A common hypothesis in adaptive AI research is that minor differences in people's predisposition to trust can significantly impact their likelihood of complying with recommendations from the AI. Predisposition to trust is often measured with self-report inventories that are administered before interactions. We benchmark a popular measure of this kind against behavioral predictors of compliance. We find that the inventory is a less effective predictor of compliance than the behavioral measures in datasets taken from three previous research projects. This suggests a general property that individual differences in initial behavior are more predictive than differences in self-reported trust attitudes. This result also shows a potential for easily accessible behavioral measures to provide an AI with more accurate models without the use of (often costly) survey instruments. 

\end{abstract}
\section{Introduction}
The capability to accurately model the attitudes that people will have towards an AI, such as trust, is critically important to the design of successful human-AI interactions. Lee and See \cite{lee2004trust} argue in their seminal work on trust in automation that minor differences between people in their predispositions to trust can result in significantly different interactions with an autonomous system. For example, a person with a high predisposition to trust may take a chance and comply with the system's recommendation, while someone not so predisposed would not. The hypothesis goes that, if this autonomous system is sufficiently reliable, then the compliant teammate will most likely experience a positive outcome and, given the assumption that positive outcomes are predictive of increased trust, such an experience should in turn lead to more compliance, thus the act of complying will initiate a feedback loop in which trust begets trust. A person low in predisposition to trust, however, will be less likely to comply and subsequently less likely to enter the positive feedback loop. 

Adaptive AI with the capability of modeling and integrating a person's predispositions, trust or otherwise, into its decision making may be able to avoid the negative effects of miscalibrated attitudes that people hold towards it. In the example of the non-compliant teammate, such an AI may attach a more in-depth explanation and justification with its recommendation, aimed at overcoming this person's deficient predisposition to trust. Such personalization of the recommendation necessitates continuous measuring and modeling of the trust-compliance relationship. Although Lee and See provide a model of the trust implications of a human and automated system team, they do not offer a measurement method.

Researchers have committed significant resources to study Lee and See's hypothesized relationship between predisposition to trust and trust in automation, including how to measure and adapt accordingly \cite{hoff2015trust}. A common practice is to measure a person's predisposition to trust using a general-purpose scale and then correlate that measure with human behavior in an interaction with an automated system. A popular example of such a scale is the 12-item \textit{Disposition to Trust} inventory (DTI) \cite{mcknight2002developing}. This measure attempts to capture a person's general likelihood of depending on others and is hypothesized to influence beliefs and intentions towards not only people but also other systems that are subject to a person's trust. There is evidence of this measure having (e.g. \cite{rossi2018impact,aliasghari2021effect}) and not having (e.g. \cite{wang2016impact,rossi2020evaluating}) predictive value in interactions with both virtual and physical robots. 

Understanding how individual characteristics, like predisposition to trust, impact compliance is more critical now than ever. Researchers, technologists, and funding agencies are increasingly pursuing automated systems capable of rich interactions with humans (e.g. \cite{amershi2019guidelines,elliot,seeber2020machines,shneiderman2020human}). Despite these efforts, AI agents will continue to make mistakes due to the complexity and inherent uncertainty of the environments in which they operate \cite{amershi2019guidelines}. One solution for mitigating the damage done by these mistakes, which can lead to better-calibrated trust, is for an agent to ensure that its reasoning is communicated in a way to meet the varying information needs and preferences of its human teammates. Sufficient transparency in communication will allow people to know when they should comply with an AI's recommendation and when they should not, i.e. develop better-calibrated trust, which is an important precursor to good compliance behavior in human-AI interactions \cite{lee2004trust,barnes2021human,wang2016impact,rossi2018impact,aliasghari2021effect}. 

On the other hand, if an agent is \emph{insensitive} to people's different predispositions to trust, it will use the same communication strategy for all of them. Given the ubiquity of individual differences, this one-size-fits-all strategy is sure to lead to unwarranted misuse or disuse of the agent \cite{parasuraman1997humans}. This makes predicting trust as early as possible---perhaps even as a trait, as DTI aims to do---all the more important. If, for example, a robot can form a model of a new human teammate's predisposition to trust before an interaction even begins, it can get a head start on personalizing its communication to maximize transparency, trust calibration, and appropriate compliance. This reality points to an acute need for compliance prediction benchmarks that are not only accurate but which agents can acquire as early as possible without sacrificing predictive value. 

We benchmark DTI against a set of four simple behavioral measures: whether or not a person follows a robot's first recommendation, its recommendation immediately after its first mistake, all its recommendations through that mistake, and its recommendations during the course of an early mission. We compare these measures in terms of their ability to predict objective (behavioral) outcomes, using data from three separate experiments collected to study human-robot interactions. DTI explains less variance in (and thus has less predictive power of) the performance of the human-AI teams than the behavioral measures across all three datasets, and contrary to the hypothesized effect of \cite{mcknight2002developing}, does not predict initial or overall compliance. In two of the three datasets, DTI does not even manage to explain a significant amount of variance in the models, meaning that it does not have predictive value as modeled. In the third dataset, it is inversely correlated with compliance. That is, whereas higher DTI scores are hypothesized to predict more compliance, they actually predict the opposite. Given that administering a survey like DTI is impractical (and sometimes impossible) in many human-AI interaction domains, it is encouraging that these early behavioral measures (which are typically readily observable by the agent) can provide such predictive power. In fact, assuming that the hypothesized relationship between compliance and trust is accurate, they appear to be more accurate reflections of a person's true predisposition to trust than a measure like DTI, which is based on self-report.

We believe that this relatively simple finding will provide valuable input to models across a diverse set of human-AI interaction domains. There are, admittedly, many existing subjective measures of trust intended for predicting compliance---but administering said measures is often burdensome and they are not always as predictive as anticipated. Likewise, there are many computational models of trust used for predicting compliance, but they are mostly domain specific and almost always challenging to implement. On the other hand, our behavioral method is simple to assess, easy to implement, and, most importantly, predictive of compliance.  Moreover, it holds the potential to aid in developing more robust predictors of compliance, identify when compliance interventions may be necessary, and serve as a prototype for other similar benchmarks. We argue that identifying and developing easy-to-assess measures, such as the behavioral measures of trust presented here, are critical steps toward developing robust, adaptive AI that are capable of personalized interventions. 

\section{Related Work}
``Trust'' is often used as an umbrella term that covers a variety of human factors that influence compliance (see \cite{hancock2011meta} for a meta-analysis that catalogs such factors). Measuring and modeling some of these specific trust factors is a common practice in AI research. Beyond the aforementioned DTI, researchers have developed or adapted many other instruments in service of understanding them, typically in isolation from one another. The adoption of these instruments by AI researchers generally reflects the nuances of each target factor. Researchers can use these broad measures to establish a hypothetical baseline for each person, while a more specific measure may be adopted for other purposes, just as researchers can use DTI to provide a baseline on a particular person's predisposition to trust. It is beyond the scope of this investigation to examine all of these factors and instruments, so we focus on those related to a person's a priori inclination to trust in an autonomous system. 

Whereas DTI seeks to identify human teammates' inherent levels of trust, the \emph{propensity to trust inventory} seeks to identify whether they are inclined to trust \cite{ashleigh2012new}. The hypotheses related to its implementation are similar: people with a greater propensity to trust will be more likely to comply with recommendations from an automated system than those with a lower one. Researchers have deployed this measure as a means of controlling for individual variance in responses to social engineering by a robot \cite{aroyo2018trust,aliasghari2021different} or to understand the relationship between attention control and trust \cite{textor2021paying}.

The context of an interaction may alter how a person responds to an automated system. With this in mind, researchers have developed and adopted trust measures for numerous use cases. At one level of abstraction are scales developed for assessing trust in particular types of automation, like the human-robot interaction trust scale, which (as its name suggests) is for assessing the level of trust a person has in robots \cite{yagoda2012you} or the human-robot trust scale which was developed to assess changes in a person's trust in a robot \cite{schaefer2013perception}. Researchers have further refined scales such as this, DTI, and others for specific use cases. The Social Service Robot Interaction Trust scale, for example, was designed with a specific focus on the service context of human-robot interactions \cite{chi2021developing}. 

Finally, there are computational methods to track the evolution of trust during extended interactions with robots. These methods, such as the probabilistic model developed in \cite{xu2015optimo} or the Markovian model introduced in \cite{pynadath2019markovian}, allow a robot to dynamically adapt its representation of the trust that a person places in it. Typically, such methods are seeded with something like the results of a trust inventory like the human-robot trust scale but obviate the need for repeated measures. 

\section{Measuring Disposition to Trust}

McKnight et al. \cite{mcknight2002developing} developed DTI (presented in Appendix A as Figure \ref{fig:DTIbox}) as part of a larger construct, specifically aimed at e-commerce, for predicting the outcome of an initial interaction with a vendor. It has four sub-measures of its own: benevolence, integrity, competence, and trusting stance (each has three items, ordered respectively). Typically, the assessment asks people to indicate how much they agree with each statement using a response scale. These responses are averaged and correlated with the outcome measure of interest, such as compliance. The basic assumption is that if a person scores high on DTI, then they will be more likely to trust (share private information with, make a purchase from, take advice from, etc.) a given vendor which will result in them complying with a request. 

The tacit assumption in adapting a measure of trust, such as DTI, to human-AI interactions or similar settings is that it remains reliable and that the construct validity observed in the development of the measure will hold in the new domain. At the time of writing, the article introducing this measure had over 6000 citations including numerous applications in human-AI interactions (e.g. \cite{wang2016impact,wang2016trust,rossi2018impact,wang2018my,lutz2020robot,rossi2020evaluating,tauchert2019following,aliasghari2021effect,aliasghari2021different,explainableRL}). No publications were found that validated the scale for use in human-AI interactions settings. 

Additionally, it is worth noting that the original authors did not validate DTI against actual trusting behavior. As they point out, common trust-related behaviors observed in e-commerce are sharing personal information, making a purchase, or acting on information provided via a website. None of these were utilized by the authors to validate their measure because of cost. Instead, study participants were asked to indicate their likelihood of performing different trusting behaviors, which the authors justified from prior research that suggested responses to such measures do not differ meaningfully from actual behavior (e.g. \cite{venkatesh2000determinants}). Summarily, this inventory was designed to study purchase intentions in e-commerce settings, not compliance potential in human-AI interactions. 

These two details, that the measure was developed for a different setting and as a means of predicting intentions rather than behaviors, suggest that the scale may be a poor fit for building a model for human-AI interactions compliance. There are other measures available, like the propensity to trust measure used by \cite{wang2016impact} and measures recently developed for predicting trust in automation \cite{jessup2019measurement,merritt2014continuous}, but their adoption is limited and, again, they were not designed specifically for human-AI interactions settings either. There are still others developed for specific types of attitudes, like the negative attitudes towards robots scale \cite{nomura2006measurement}, but, again, they are not designed specifically for predicting compliance. This all points to the need for other means of informing a model capable of predicting compliance behavior in human-AI interactions. 

\section{A Behavioral Predictor of Compliance}

\textit{``The best predictor of future behavior is past behavior.''}

\hspace*{\fill} - \sout{Walter Mischel}

\hspace*{\fill} - \sout{B.F. Skinner}

\hspace*{\fill} - \sout{Mark Twain}

As the strikeouts suggest, this maxim has been floating around in the study of human behavior for a significant amount of time. Despite its resiliency, the maxim is still subject to the same critique as all maxims, does it \textit{really} generalize? The answer appears to be yes, given a small set of caveats: a short time horizon, similarity in the decision scenario, and stability in the decision maker \cite{mischel2013personality}. We posit that the past behavior maxim applies to human-AI interactions and that the best way to judge future compliance with recommendations from an AI counterpart is to observe one (or a small number of) interaction and extrapolate.

Past behavior is more than a simple measure of habit. It reflects a suite of factors that weigh on decision making in a reliable fashion. This drives the decision-maker toward self-consistency and is, at least in part, what gives past behavior its predictive power \cite{ajzen1991theory,ouellette1998habit}. It also suggests that when past behavior is able to explain residual error left by a theoretical construct that the construct is not capturing some important predictive factor(s). Simply put, the construct is insufficient. This alone does not mean that the construct lacks explanatory value. It is possible (likely) that trust, for example, is only a small part of compliance and past behavior is predictive of future compliance because it captures some heretofore unidentified factors. 

We do not believe that this is the case with measures of trust used to predict future compliance in many human-AI interaction experiments. This is because these trust measures, like DTI, fail to account for a significant amount of variance in regression models. In other words, they do not have predictive power. Moreover, our alternative predictor, prior behavior, does. It is also possible to test whether DTI or past behavior can account for variance in the data that the other did not. Testing whether past behavior can explain away residual error left by DTI in the datasets is straightforward: simply add a control for past behavior to a statistical model that includes DTI and observe whether residual error significantly goes down. Moreover, the same basic concept can be applied for determining if DTI explains a meaningful amount of variance in behavioral data by comparing two models, one with DTI as a regressor and one without (that is, with just the experimental treatment controls). 

\section{Empirical Strategy}
We analyzed three datasets, described in Appendix B, to test whether DTI and past behavior are capable of predicting compliance, i.e. future behavior (FB), in human-AI interactions. We introduce four different measures of past behavior. The simplest measure of past behavior is the participants' first compliance choice (FC) to follow or ignore a robot's recommendations during an experimental session. Studies 1 and 2 were comprised of multiple missions (three and eight, respectively). For these two data sets, we compute participants' mission 1 compliance (M1C) as an extended measure of past behavior. The third study only consisted of one long mission, however, every participant in this study followed the same trajectory with the robot making the same set of mistakes. Thus, we use participants' compliance on the choice after the first mistake (AFM) made by the robot and average compliance through AFM (AC-AFM). Future behavior, the dependent variable for the various models, is the percentage of times in the remaining interactions that a participant complied with the robot's recommendations. Thus, the models with FC as an independent variable and the models presented for comparing them use a different dependent variable (average compliance for buildings 2:\textit{n}) than the models with M1C (average compliance after mission 1) or AFM, and AC-AFM (average compliance for buildings \textit{mistake}+2:\textit{n}) as independent variables. Note that two of the data sets included multiple missions, meaning that there was a break between sets of buildings. Our analyses ignore this fact and treat all buildings the same.\footnote{Controlling for mission did not meaningfully change the interpretation of the results.} We describe the data and models in full detail in Appendix B. 

\section{Results}

As noted, we rely on regression analyses to model the data. Each regression model that we report for a given use case is presented as a column in a table in the appendix. In the tables that report regression models, the first column is always the reference (null) model. These are models that do not include DTI or prior behavior measures, simply experimental treatment controls. We do not report values for the treatment conditions as these are reported in the original papers. Coefficients for continuous variables, such as a participant's averaged DTI response, should be interpreted as the expected change in the dependent variable, all else being equal, if the variable in question increased by one. For indicator variables, such as whether a participant complied during their first interaction with the robot, the coefficient value indicates the average change, all else being equal, in the dependent variable when the independent variable goes from zero to one. 

\subsection{Study 1}
The disposition to trust measure failed to explain a meaningful amount of variance in the data from Study 1. When predicting the compliance percentage for buildings 2:24, the penalty of having an additional regressor in the model meant that the adjusted $R^2$ was no better than that of the reference model (see columns (1) and (2) of table \ref{study1FCReg}; all tables are in Appendix C). Unsurprisingly, the \textit{F}-test comparing these two models rejects the alternative hypothesis ($F=0.832,\, p=0.363$). However, adding the FC measure (see column (3) of table \ref{study1FCReg}) did result in a model that accounted for more variance than the reference model ($F=11.032,\, p=0.001$). The complete model with both measures is better than the DTI ($F=10.575,\, p=0.001$) model but not the FC ($F=0.449,\, p=0.504$).

Shifting from using one interaction to predict the next 23 to using the entirety of mission one compliance decisions to predict those of the next two missions (16 interactions) resulted in generally better model fits. In effect, this is a sanity check because we are using more of a given participant's past behavior to predict less of their future behavior. The resulting models, reported in table \ref{study1M1CReg} and compared in table \ref{study1M1CComp}, reflect those for the FC measure. The M1C measure explained a meaningful amount of variance but DTI did not. Together, these results support the hypothesis that prior behavior was a better predictor of future compliance than DTI for study 1. 

\subsection{Study 2}
The results for the FC measure from Study 2 replicated those of Study 1, as reported in table \ref{study2FCReg}; however, it should be noted that the model fits were generally poor (see the $R^2$ values in table \ref{study2FCReg}). The DTI model (2) was no better than (1) while (3) managed to explain significantly more variance in the parameters. Adding the DTI measure to (3), again, did not improve the model's performance. This means we can draw the same conclusion from Study 2 as we did from Study 1: FC explains variance in compliance behavior not captured by DTI. The M1C models were also generally poor fits (see the $R^2$ values in table \ref{study2M1CReg}). Moreover, they did not replicate the findings of the M1C models from Study 1. It is unclear if this is an artifact of the difference in study designs, the low $n$, or actually reflective of the quality or generalizability of the measure. Although we cannot conclusively state that prior behavior predicted future behavior during study 2, the data do suggest that prior behavior was a better predictor than DTI. 

\subsection{Study 3}
Study 3 offers more intriguing results than the two earlier experiments. DTI did predict future behavior, however, it was negatively correlated. In other words, participants that scored higher on DTI were less likely to comply with the robot's recommendations over the course of the experiment (see column (2) in tables \ref{study3FCReg}, \ref{study3AFMReg}, and \ref{study3ACAFMReg}). The model fits suggest that a 1 unit higher score on the DTI measure would predict about 5\% \emph{less} compliance, roughly the same as ignoring the robot's advice for two buildings more than the average participant. This effect reverberates in the \textit{F} test: Model (2) accounts for significantly more variance than Model (1), thus we reject the null hypothesis when predicting compliance for buildings 2:45 ($F=8.885,\, p=0.003$) as well as the shorter horizon 9:45 ($F=9.732,\, p=0.002$).

The models also suggest that the FC, AFM, and AC-AFM measures were effective means of predicting future compliance behavior. Participants who complied during the first interaction complied nearly 12\% more for the remaining missions (column (3) of table \ref{study3FCReg}). A similar effect was observed when using the first compliance decision after the robot's first mistake (AFM, in which case compliant participants were nearly 19\% more compliant during remaining missions; column (3) of table \ref{study3AFMReg}) and when using the average compliance through the first compliance decision after the robot's first mistake (AC-AFM, in which case the model suggests a roughly 31\% difference in future compliance between participants that were perfectly non-compliant versus perfectly compliant; column (3) of table \ref{study3ACAFMReg}). 

In the case of the models including both DTI and past behavioral measures, for Study 3, we actually reject the null hypothesis that including DTI does not contribute to the models with just behavioral measures (AF: $F=8.985,\, p=0.003$, AFM:$F=8.267,\, p=0.005$, AC-AFM: $F=9.046,\, p=0.003$). Importantly, the predicted effect of DTI remains negatively correlated with compliance and with about the same impact. Summarily, the models we present over these three studies offer two insights: First, past compliance behavior, even in its simplest form, is a more reliable, consistent predictor of future compliance, and second, DTI is not as reliable and may even hold an inverse correlation from what is hypothesized in the literature. 

\subsection{An Alternative Explanation}
McKnight et al. \cite{mcknight2002developing} specifically state that DTI should be positively correlated with early compliance, not necessarily overall compliance. It is possible that DTI does actually predict a participant's first compliance choice (FC). This means that, rather than comparing FC and DTI, DTI should be employed to predict FC. We tested this hypothesis using multiple linear regression for each study and report the results in table \ref{DTIhypTab}. We also included treatment condition controls in each model; as in the other tables, their presence is indicated by the checkmarks and the associated coefficient values and intercepts are redacted as a simplification. Summarily, DTI did not significantly predict participants' first compliance choice in any of the studies. We interpret this as strong evidence in support of the null hypothesis that DTI is not correlated with early compliance choices. We also expanded the definition of early compliance to include the first five compliance choices; however, the result persisted so we did not include the results as an additional table.  

\section{Discussion}
Our review and empirical evaluation of past research that employed the Disposition to Trust Inventory led us to conclude that DTI is not an accurate predictor of compliance in human-AI interactions. Data from three prior experiments support this conclusion. Additionally, the basic method that we employ can serve as a prototype for evaluating other instruments.

Across the studies from which we drew data, the robots did significantly better than chance at predicting when protective gear is needed (e.g., even the low-ability robot is 80\% reliable). This means that compliance was highly correlated with mission success. Just as early compliance behavior predicts later compliance, the data also suggest that it can predict success. Drawing conclusions from this, of course, is problematic because it is an artifact of the correlation. We believe that a more robust study of early behavior and its relationship to later compliance is warranted. A deep understanding of this relationship may prove useful in the design of compliance interventions, even if it is not applicable in every situation (one-shot interactions, obviously, cannot benefit from early behavior insights).

Had DTI accounted for more significant variance in the models, we could have tested whether it captured variance explained by the future behavior measures and vice versa. Unfortunately, the datasets at our disposal simply did not allow for such analyses. Larger datasets, possibly from different settings in which DTI has more predictive value, may allow researchers to differentiate the factors captured by each measure that weigh on compliance.

It is worth acknowledging that trust and compliance are not perfectly aligned concepts. In the HCI and HRI spaces, it is common to reference Lee and See's definition of trust: \textit{the attitude that an agent will help achieve an individual's goals in a situation characterized by uncertainty and vulnerability} (p. 51 \cite{lee2004trust}). This simple definition belies a very complex phenomenon of human cognition: the attitudes that people form about subjects and their properties. On the other hand, compliance is relatively simple: did a person (appropriately) follow the advice given by an automated system? The hypothesis present in much of the literature on trust and compliance is that humans maintain a specific, identifiable, attitude (trust) that has broad applications, including predicting compliance, with respect to automated systems and their myriad properties. Moreover, the hypothesis posits that this attitude is easily measured using short batteries of questions, such as DTI. Generally speaking, the attitudes that humans maintain are notoriously hard to study---we argue that this feature of attitudes leaves such approaches prone to failure, particularly when the behavior of interest is not directly linked to a given measure. 

In the experiments from which we drew data, the robot is the assumed subject, and its reliability in predicting dangers is the implied property of interest. In order for the standard hypothesis to find support, the following items, minimally, need to hold:
\begin{itemize}
    \item The disposition to trust hypothesis that, to some measurable degree, trust is a trait-like aspect of cognition.
    \item Disposition to trust is accurately measured via an inventory such as DTI.
    \item Disposition to trust universally maps to subjects and their properties, i.e. there are no exceptional cases or variables that mediate or moderate the relationship.
    \item Compliance is positively correlated with disposition to trust.
\end{itemize}
If any of these do not hold, we would expect DTI to underperform or even fail as a predictor of compliance, which is exactly what we observed. Unfortunately, the available data do not allow us to further investigate in what sense we reject the standard hypothesis. 

Lee and See point out that trust is a multidimensional construct and even state that, ``There does not seem to be a reliable relationship between global and specific trust...Developing a high degree of functional specificity in the trust of automation may require specific information for each level of detail of the automation.'' (p. 58 \cite{lee2004trust}). This insight alone, which is supported by extensive research in cognitive science, seems to undermine the first, second, and third items above. These three items are further eroded with the acknowledgment that the thoughts and attitudes which people maintain towards different subjects have long been topics of scientific inquiry that are known for their recalcitrance to consensus. Notable philosophical efforts to tame them dot history. From Aristotle's efforts in \textit{On Interpretation} \cite{stevenson_2009} to Daniel Dennett's \textit{The Intentional Stance} \cite{dennett1987intentional} and Ruth Millikan's biosemantics \cite{millikan1989biosemantics} there are many laudable attempts. Nevertheless, open questions remain about what it means to have an attitude toward a subject. Modern psychology is similarly plagued: although Brentano, who heavily influence the German tradition of empirical psychology from Sigmund Freud on through Bertrand Russell, attempted to corral the subject with his work on intentionality \cite{brentano2012psychology}, it is still a hotly debated matter \cite{chater2022paradox}.

The last item, that compliance is positively correlated with trust, is similarly fraught. The recent pandemic serves as a convenient case study: Trusting stances were found to have a significant positive correlation with European citizens' compliance with government policies during the early pandemic \cite{bargain2020trust}, however, in Singapore, despite general trust in the government, compliance was poor even with efforts from the government to change citizens' risk perceptions \cite{wong2020paradox}. Together, these findings call into question the reliability of compliance's correlation with trust. Summarily, with the addition of our analyses, we believe that the standard hypothesis fails on multiple fronts: it is simultaneously too broad given its reliance on a trust-like trait, and too specific in its ignorance of important confounds. 

Nevertheless, there may still be value in measures such as DTI, particularly when they are shown to have predictive value (no matter the directionality of the correlation). If a measure reliably accounts for variance in human behavior, it has value in developing automated systems since explaining variance is core to predicting choices. As far as we are aware, our analyses do not refute the findings of any human-AI interaction paper. Our key takeaway is that we should not only vet our measures with thorough literature reviews but also validate them for the specific modeling that is required for human-AI interaction domains.

Our overall recommendations for the community are:
\begin{itemize}
    \item \textbf{Validate psychometric measures adopted from other disciplines for your specific model and outcome measure.} Additionally, as pointed out by \cite{schrum2020four}, using a single item (or some subset of items) from a Likert scale undermines its original reliability and validation. 
    \item \textbf{Consider alternative, simple measures that may already exist in your data.} These measures do not necessarily need to replace others but can serve as valuable companions and provide more robust inference.
    \item \textbf{Evaluate the usefulness of new measures against alternatives.} The statistical work put into understanding the ways in which different measures account for variance in data not only ensures a more rigorous examination of these alternate causes but also can produce new insights and drive new modeling formulations.
\end{itemize}
It would be uncanny if the effects that we observed in the data are unique to DTI, trust, and compliance. It seems highly probable that similar effects exist for other mental states that humans maintain towards automated systems, such as anger, belief, fear, hope, etc., and their affiliated outcomes. For example, extensive research suggests that negative attitudes towards robots can be highly predictive of interactions and outcomes \cite{nomura2006measurement,nomura2006experimental}. We would hypothesize that, just as early compliance is a better predictor of later compliance than DTI, there are behavioral measures that can outperform and/or are less intrusive than the negative attitudes towards robots instruments. Identifying, testing, and validating such measures, we believe, will prove a critical step towards improving relationships between humans and all types of automated systems, from recommender to robot. 

\section{Conclusion}
Inventories measuring different psychological constructs are popular for informing models in human-AI interaction and are frequently misused \cite{schrum2020four}. In this paper, we evaluated the usage of one such measure. This measure was used in prior research in hopes that it would predict important human behaviors, like compliance, and overall outcomes, like task performance. The measure failed in two experiments and had the opposite correlation predicted by the measures' creators in a third. We found that a simple behavioral measure was a better predictor of future compliance and would thus be a more informative input to a model of compliance reasoning. The methodology used here to evaluate the two measures is a straightforward prototype for other researchers to use in evaluating alternate measures that can serve as input into models in human-AI interaction.
%
%
%
 \bibliographystyle{splncs04}
 \bibliography{bib.bib}
\newpage
\begin{subappendices}
\renewcommand{\thesection}{\Alph{section}}

\section{Figures}

\begin{figure}
\caption{Disposition to Trust Inventory Items}
\label{fig:DTIbox}
\fbox{
    \begin{minipage}{0.90\columnwidth}
        \begin{description}
        \item[Benevolence]\ 
            \begin{enumerate}
            \item In general, people really do care about the well-being of others.
            \item The typical person is sincerely concerned about the problems of others.
            \item Most of the time, people care enough to try to be helpful, rather than just looking out \\for themselves.
            \end{enumerate}
            
        \item[Integrity]\ 
            \begin{enumerate}
            \item In general, most folks keep their promises.
            \item I think people generally try to back up their words with their actions.
            \item Most people are honest in their dealings with others.
            \end{enumerate}
            
        \item[Competence]\ 
            \begin{enumerate}
            \item I believe that most professional people do a very good job at their work.
            \item Most professionals are very knowledgeable in their chosen field.
            \item A large majority of professional people are competent in their area of expertise.
            \end{enumerate}
            
        \item[Trusting Stance]\ 
            \begin{enumerate}
            \item I usually trust people until they give me a reason not to trust them.
            \item I generally give people the benefit of the doubt when I first meet them.
            \item My typical approach is to trust new acquaintances until they prove I should not trust them.
            \end{enumerate}
            
        \end{description}
    \end{minipage}}
\end{figure}
\newpage
\begin{figure*}
\label{fig:Compliance}
\includegraphics[width=0.95\textwidth]{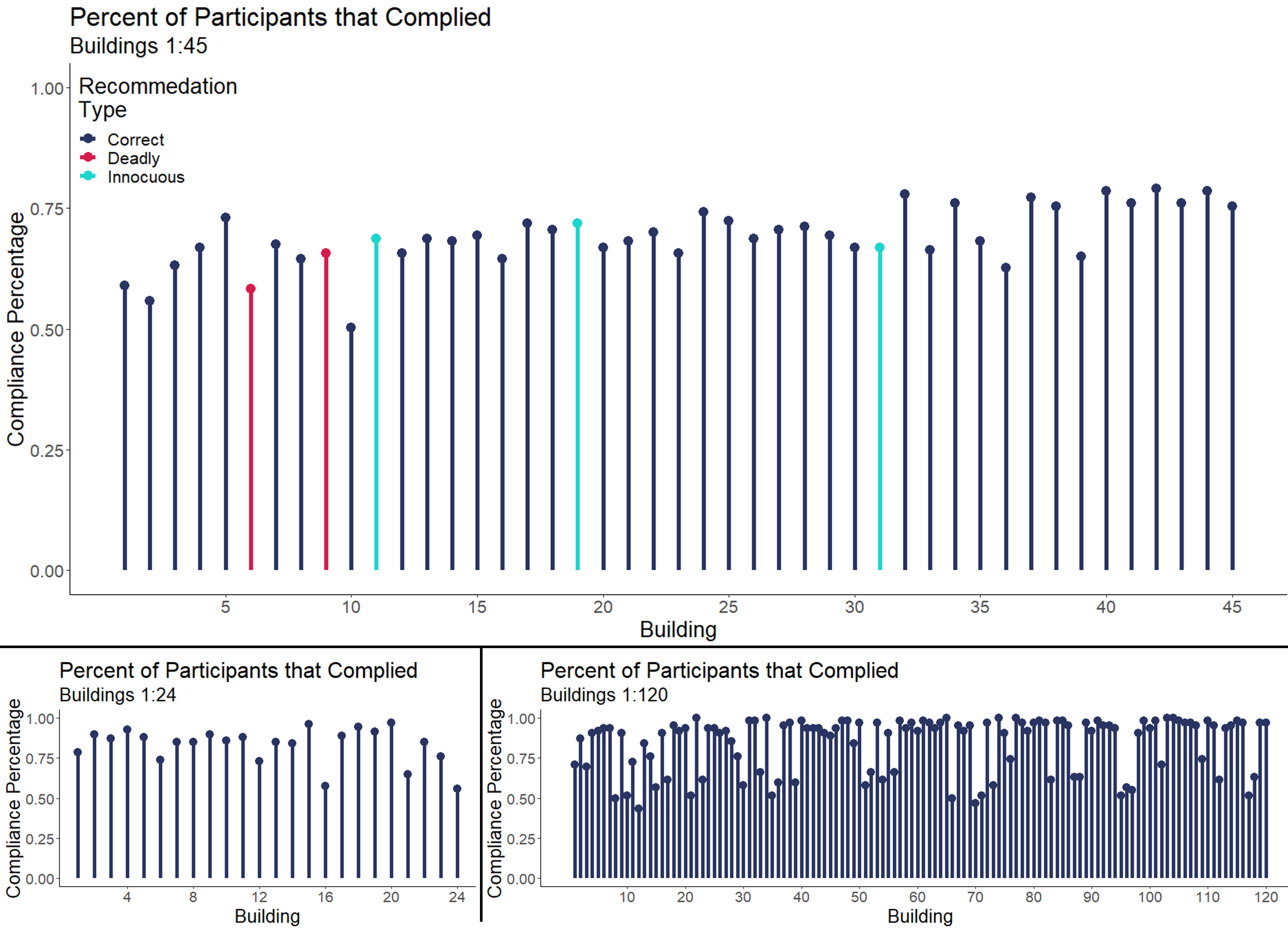}
\caption{The robot only communicated differently in study 3---it always made the same mistakes, in the same order. These features mean that the data lend themselves to concise, straightforward visual representation, which we present in the top panel. Note that the colors in this panel indicate the robot's recommendation type. Specifically, the robot made two deadly and three innocuous mistakes which are highlighted above. Overall, there is a trend of increasing compliance. The figures for studies 1 and 2, which are respectively the bottom left and right panels, only indicate compliance percentage, not the type of recommendations made by the robot.}
\end{figure*}
\newpage
\section{Data and Models}
\subsection{Models}
We used linear regression to model and test the predictive value of the various behavioral measures. The basic approach is to fit a reference model (\textbf{Model 1}) in which the outcome measure, future behavior, is predicted by the treatment conditions alone, such as:
\begin{equation}
    Y_{iFB} = \beta_0 + \beta_\mathrm{Treat} X_\mathrm{iTreat} + \epsilon_i
\end{equation}
We use this general form for the reference (null) model across the experiments. Note that there are different \textit{FB} measures; which one is used in a given set of models depends on the independent variable in question. A given reference and test model, however, always use the same dependent variable. 

\textbf{Model 2}: The first category of test models incorporate participants' DTI score as an independent variable:
\begin{equation}
    Y_{iFB} = \beta_0 + \beta_\mathrm{Treat} X_\mathrm{iTreat} + \beta_\mathrm{DTI} X_\mathrm{iDTI} + \epsilon_i
\end{equation}
Note that the reference model is nested within this model. In other words, equation (2) represents the alternative hypothesis that adding the predictor variable to the model will result in accounting for more variance, or a significant improvement in model performance. Comparing the two models is as simple as applying an \textit{F}-test, which in this case tells us whether the more complex model results in a statistically different residual sum of squares value. If it does, we can reject the null hypothesis, i.e. that the reference model is sufficient, in favor of the alternative hypothesis, or that adding the additional predictor variable(s) was warranted. The \textit{F}-test, in this instance, can be thought of as allowing us to investigate the utility of adding DTI or other measures to the model.\footnote{A likelihood ratio test is another method of comparing such models and will produce similar insights.} When this test returns a p-value less than 0.05, we can conclude that the residual sum of squares values for the two models are significantly different at the level of $\alpha = 0.05$, or in other words, the new explanatory variable is warranted as it significantly increases the amount of variance explained in the model (i.e. RSS is lower).

\textbf{Model 3}: The next category of test models incorporate participants' past behavior as independent variables, for example, the model:
\begin{equation}
    Y_{iFB} = \beta_0 + \beta_\mathrm{Treat} X_\mathrm{iTreat} + \beta_\mathrm{FC} X_\mathrm{iFC} + \epsilon_i
\end{equation}
This uses participants' first choice to predict their compliance for all of the remaining choices that they faced. Again, this has the reference model embedded in it, and a simple \textit{F}-test will reveal the utility of the FC predictor. We construct similar models for M1C, AFM, and AC-AFM (but using appropriate FB measures). 

\textbf{Model 4}: Since it is likely that DTI and the past behavior measures are accounting for different variance in the models, directly comparing equations (2) and (3) via $R^2$ is not entirely informative. Thus, we introduce a third category of test models in which both DTI and a past behavior measure are included. In the case of FC, the model is:
\begin{equation}
    Y_{iFB} = \beta_0 + \beta_\mathrm{Treat} X_\mathrm{iTreat} + \beta_\mathrm{DTI} X_\mathrm{iDTI} + \beta_\mathrm{FC} X_\mathrm{iFC} + \epsilon_i
\end{equation}
These models facilitate assessing whether the added complexity of including both DTI and the past behavior measure is warranted, again using \textit{F}-tests. Finally, for readability, we refer to relevant statistics within the text of the manuscript but place regression and other tables for each set of models and their associated tests in the appendix. These tables include, in the same order as above, models that facilitate comparing DTI and the behavioral measures. Each regression table is followed by a table presenting the results of the relevant \textit{F}-tests.

\subsection{Data}
We used data from experiments conducted as part of a long-term research project on explainability and AI. Participants of these experiments team with a simulated robot during reconnaissance missions. The missions involve entering buildings to determine whether threats are present. The robot goes first and is equipped with a camera, microphone, and sensors for nuclear, biological, and chemical threats. These sensors are not perfectly reliable. Based on the data that it collects using its sensors, the robot makes a recommendation to the participant about putting on protective gear. The participant then makes a choice about wearing the gear, i.e., whether or not to comply with the robot's recommendation. When participants wear the gear, it always neutralizes any threat. If they do not wear it and encounter a threat, they die in the virtual world, but in reality, incur a prohibitive time penalty. Finally, participants incur a slight time delay (much smaller than that for dying in the virtual world) when equipping the gear. 

In all three studies, the robot based its recommendations on the noisy sensor readings as input to a policy computed through either Partially Observable Markov Decision Processes (POMDPs) \cite{Kaelbling1998} or model-free reinforcement learning (RL) \cite{kaelbling1996reinforcement,sutton2018reinforcement} using the reward signal of the time cost and deaths incurred. The robot performed significantly better than chance across the studies. This means that compliance from the participants was highly correlated with making the normative choice, i.e., wearing protective equipment at the right time. 

Participants of all three studies completed the 12-item DTI before starting their assigned mission(s). Gross experimental details and the results of these conditions are reported in the original publications. As such, we do not replicate those findings here for brevity's sake. Do note, however, that the $n$'s we report may vary from the original papers because of incomplete observations (some participants chose to not complete the DTI). 

Study 1 participants ($n=198$, Amazon Mechanical Turk) completed three missions, each with eight buildings \cite{wang2016impact}. They were randomly paired with one of two POMDP-based robot types: a high-ability robot that was never wrong or a low-ability robot that made mistakes 20\% of the time (or was 80\% reliable). Both types of robots were crossed with four different recommendation explanation conditions: none, confidence level, sensor readings version 1, and sensor readings version 2. This experiment was fully between subjects, meaning that each participant interacted with only one robot type and received only one type of explanation throughout the missions. The coefficient $\beta_\mathrm{Treat}$ in the models for Study 1 captures which information condition participants experienced. First compliance choice (FC) takes 1 if participants heeded the robot's recommendation for the first building and 0 if not. Mission 1 compliance (M1C), on the other hand, is the fraction of times that a participant complied with the robot's recommendations during the first mission. The compliance future behavior (FB) measure associated with FC for study one is thus the fraction of times a participant complied for the remaining 23 buildings. Similarly, for M1C it is the fraction of times that a participant complied with the robot's advice during missions two and three. It should be noted that participants were not told whether they were interacting with the same robot across missions; instead, the robot started each mission as if it had not previously interacted with a participant. 

Study 2 participants ($n=53$, cadets at West Point Academy) completed eight missions, each with a different POMDP-based robot \cite{wang2018my}. In each mission, the human-robot team carried out a reconnaissance task of 15 buildings. The mission order was fixed (i.e., always searched the buildings in the same order and across missions), but the robot order was randomized. The $2\times2\times2$ design crossed robot acknowledgment of mistakes (none/acknowledge), recommendation explanation (none/confidence), and embodiment (robot-like/doglike). Unlike Study 1, participants interacted with different robots during each mission. Nevertheless, to demonstrate the robustness of the simple behavioral measures, we rely on the same first compliance choice (FC) as Study 1 and a similar mission 1 compliance (M1C). The compliance measures, obviously, cover a longer horizon: 119 missions and 105 missions, respectively. The $\beta_\mathrm{Treat}$ of the models for Study 2 data captures the robot type of \textit{the first mission}. It is possible that ordering of robot advisors mattered; however, the data are insufficient to specify a hierarchical model that would uncover such a feature. 

Study 3 participants ($n=148$, Amazon Mechanical Turk) completed one mission that covered 45 buildings with an RL-based (RL: reinforcement learning) robot in a fully between design \cite{explainableRL,gurney2022measuring,gurney2023my}. The treatment conditions held the robot's ability constant but varied how it explained its recommendations: no explanation, explanation of decision, explanation of decision and learning. Again, the first compliance choice (FC) is the same as the previous two studies and the FC outcome measure is the compliance fraction for the remaining 44 buildings. Mission 1 compliance (M1C) is not applicable given that the entire experiment consisted of a single mission. Given that building order and robot performance were fixed across treatment conditions, however, the two additional compliance measures, choice after the first mistake (AFM) and average compliance through the first mistake (AC-AFM), become meaningful. The first mistake occurred during building six, thus participants' decision for building seven is the AFM measure and the fraction of times they complied during the first seven buildings is AC-AFM. Relatedly, the dependent variable is the fraction of times that a given participant complied during the remaining 38 buildings.

\newpage
\section{Tables}

\setlength\tabcolsep{2pt} 

\begin{table*}[!ht] \centering 
  \caption{Study 1 FC Models} 
  \label{study1FCReg} 
\begin{tabular}{@{\extracolsep{5pt}}lcccc} 
\\[-1.8ex]\hline 
\hline \\[-1.8ex] 
 & \multicolumn{4}{c}{\textit{Dependent variable: Compliance Percentage Buildings 2:24}} \\ 
\cline{2-5} 
\\[-1.8ex] & \multicolumn{4}{c}{Regression Models} \\ 
\\[-1.8ex] & (1) & (2) & (3) & (4)\\ 
\hline \\[-1.8ex] 
 DTI &  & 0.007 &  & 0.005 \\ 
  &  & (0.008) &  & (0.008) \\ 
  & & & & \\ 
 FC &  &  & 0.086$^{**}$ & 0.084$^{**}$ \\ 
  &  &  & (0.026) & (0.026) \\ 
  & & & & \\ 
 Treatment Controls & \checkmark & \checkmark & \checkmark & \checkmark \\ 
  & & & &\\ 
 Constant & 0.762$^{***}$ & 0.724$^{***}$ & 0.687$^{***}$ & 0.661$^{***}$ \\ 
  & (0.032) & (0.052) & (0.038) & (0.054) \\ 
  & & & & \\ 
\hline \\[-1.8ex] 
Observations & 198 & 198 & 198 & 198 \\ 
R$^{2}$ & 0.420 & 0.422 & 0.452 & 0.453 \\ 
Adjusted R$^{2}$ & 0.398 & 0.398 & 0.429 & 0.427 \\ 
\makecell[l]{Residual Std.\\ \hspace{1mm}Error} & \makecell[c]{0.152\\ (df = 190)} & \makecell[c]{0.152\\ (df = 189)} & \makecell[c]{0.148\\ (df = 189)} & \makecell[c]{0.148\\ (df = 188)} \\ 
\makecell[l]{F Statistic\\} & \makecell[c]{19.639$^{***}$\\(df = 7; 190)} & \makecell[c]{17.273$^{***}$\\(df = 8; 189)} & \makecell[c]{19.471$^{***}$\\(df = 8; 189)} & \makecell[c]{17.307$^{***}$\\(df = 9; 188)} \\ 
\hline 
\hline \\[-1.8ex] 
\textit{Note:}  & \multicolumn{4}{r}{$^{*}$p$<$0.05; $^{**}$p$<$0.01; $^{***}$p$<$0.001} \\ 
\end{tabular}
\end{table*} 

\begin{table}[!htbp] \centering 
  \caption{Study 1 FC Model Comparisons} 
  \label{study1FCComp} 
\begin{tabular}{@{\extracolsep{5pt}}lcccc} 
\\[-1.8ex]\hline 
\hline \\[-1.8ex] 
Statistic & \multicolumn{1}{c}{(1) Vs. (2)} & \multicolumn{1}{c}{(1) Vs. (3)} & \multicolumn{1}{c}{(2) Vs. (4)} & \multicolumn{1}{c}{(3) Vs. (4)}\\ 
\hline \\[-1.8ex] 
Sum of Sq & 0.019 & 0.242 & 0.232 & 0.010 \\ 
F & 0.832 & 11.032 & 10.575 & 0.449\\ 
Pr(\textgreater F) & 0.363 & 0.001 & 0.001 & 0.504\\ 
\hline \\[-1.8ex] 
\end{tabular} 
\end{table} 

\begin{table*}[!ht] \centering 
  \caption{Study 1 M1C Models} 
  \label{study1M1CReg} 
\begin{tabular}{@{\extracolsep{5pt}}lcccc} 
\\[-1.8ex]\hline 
\hline \\[-1.8ex] 
 & \multicolumn{4}{c}{\textit{Dependent variable: Compliance Percentage Buildings 9:24}} \\ 
\cline{2-5} 
\\[-1.8ex] & \multicolumn{4}{c}{Regression Models} \\ 
\\[-1.8ex] & (1) & (2) & (3) & (4)\\ 
\hline \\[-1.8ex] 
 DTI &  & 0.003 &  & $-$0.0004 \\ 
  &  & (0.009) &  & (0.008) \\ 
  & & & & \\ 
 M1C &  &  & 0.242$^{***}$ & 0.243$^{***}$ \\ 
  &  &  & (0.052) & (0.052) \\ 
  & & & & \\  
 Treatment Controls & \checkmark & \checkmark & \checkmark & \checkmark \\ 
  & & & &\\ 
 Constant & 0.770$^{***}$ & 0.754$^{***}$ & 0.587$^{***}$ & 0.589$^{***}$ \\ 
  & (0.035) & (0.058) & (0.052) & (0.066) \\ 
  & & & & \\
\hline \\[-1.8ex] 
Observations & 198 & 198 & 198 & 198 \\ 
R$^{2}$ & 0.451 & 0.451 & 0.508 & 0.508 \\ 
Adjusted R$^{2}$ & 0.431 & 0.428 & 0.487 & 0.484 \\ 
\makecell[l]{Residual Std.\\\hspace{1mm}Error} & \makecell[c]{0.170 \\(df = 190)} & \makecell[c]{0.170 \\(df = 189)} & \makecell[c]{0.161 \\(df = 189)} & \makecell[c]{0.161 \\(df = 188)} \\ 
\makecell[l]{F Statistic} & \makecell[c]{22.282$^{***}$ \\(df = 7; 190)} & \makecell[c]{19.421$^{***}$ \\(df = 8; 189)} & \makecell[c]{24.384$^{***}$ \\(df = 8; 189)} & \makecell[c]{21.561$^{***}$ \\(df = 9; 188)} \\ 
\hline 
\hline \\[-1.8ex] 
\textit{Note:}  & \multicolumn{4}{r}{$^{*}$p$<$0.05; $^{**}$p$<$0.01; $^{***}$p$<$0.001} \\ 
\end{tabular}
\end{table*} 

\begin{table}[!htbp] \centering 
  \caption{Study 1 M1C Model Comparisons} 
  \label{study1M1CComp} 
\begin{tabular}{@{\extracolsep{5pt}}lcccc} 
\\[-1.8ex]\hline 
\hline \\[-1.8ex] 
Statistic & \multicolumn{1}{c}{(1) Vs. (2)} & \multicolumn{1}{c}{(1) Vs. (3)} & \multicolumn{1}{c}{(2) Vs. (4)} & \multicolumn{1}{c}{(3) Vs. (4)}\\ 
\hline \\[-1.8ex] 
Sum of Sq & 0.003 & 0.568 & 0.565 & $<$0.001 \\ 
F & 0.119 & 21.923 & 21.677 & 0.002\\ 
Pr(\textgreater F) & 0.731 & $<$0.001 & $<$0.001 & 0.965\\ 
\hline \\[-1.8ex] 
\end{tabular} 
\end{table}

\begin{table*}[!ht] \centering 
  \caption{Study 2 FC Models} 
  \label{study2FCReg} 
\begin{tabular}{@{\extracolsep{5pt}}lcccc} 
\\[-1.8ex]\hline 
\hline \\[-1.8ex] 
 & \multicolumn{4}{c}{\textit{Dependent variable: Compliance Percentage Buildings 2:120}} \\ 
\cline{2-5} 
\\[-1.8ex] & \multicolumn{4}{c}{Regression Models} \\ 
\\[-1.8ex] & (1) & (2) & (3) & (4)\\
\hline \\[-1.8ex] 
 DTI &  & 0.004 &  & 0.004 \\ 
  &  & (0.011) &  & (0.011) \\ 
  & & & & \\ 
 FC &  &  & 0.043$^{*}$ & 0.043$^{*}$ \\ 
  &  &  & (0.018) & (0.018) \\ 
  & & & & \\ 
 Treatment Controls & \checkmark & \checkmark & \checkmark & \checkmark \\ 
  & & & &\\ 
\hline \\[-1.8ex] 
Observations & 49 & 49 & 49 & 49 \\ 
R$^{2}$ & 0.047 & 0.050 & 0.163 & 0.165 \\ 
Adjusted R$^{2}$ & $-$0.016 & $-$0.037 & 0.087 & 0.068 \\ 
\makecell[l]{Residual Std. \\\hspace{1mm}Error} & \makecell[c]{0.053 \\(df = 45)} & \makecell[c]{0.053 \\(df = 44)} & \makecell[c]{0.050 \\(df = 44)} & \makecell[c]{0.051 \\(df = 43)} \\ 
F Statistic & \makecell[c]{0.742 \\(df = 3; 45)} & \makecell[c]{0.574 \\(df = 4; 44)} & \makecell[c]{2.137 \\(df = 4; 44)} & \makecell[c]{1.704 \\(df = 5; 43)} \\ 
\hline 
\hline \\[-1.8ex] 
\textit{Note:}  & \multicolumn{4}{r}{$^{*}$p$<$0.05; $^{**}$p$<$0.01; $^{***}$p$<$0.001} \\ 
\end{tabular} 
\end{table*}

\begin{table}[!htbp] \centering 
  \caption{Study 2 FC Model Comparison} 
  \label{study2FCComp} 
\begin{tabular}{@{\extracolsep{5pt}}lcccc} 
\\[-1.8ex]\hline 
\hline \\[-1.8ex] 
Statistic & \multicolumn{1}{c}{(1) Vs. (2)} & \multicolumn{1}{c}{(1) Vs. (3)} & \multicolumn{1}{c}{(2) Vs. (4)} & \multicolumn{1}{c}{(3) Vs. (4)} \\ 
\hline \\[-1.8ex] 
Sum of Sq & $<$0.001 & 0.015 & 0.015 & $<$0.001 \\ 
F & 0.112 & 6.071 & 5.964 & 0.137 \\ 
Pr(\textgreater F) & 0.739 & 0.017 & 0.019 & 0.713\\ 
\hline \\[-1.8ex] 
\end{tabular} 
\end{table}

\begin{table*}[!ht] \centering 
  \caption{Study 2 M1C Models} 
  \label{study2M1CReg} 
\begin{tabular}{@{\extracolsep{5pt}}lcccc} 
\\[-1.8ex]\hline 
\hline \\[-1.8ex] 
 & \multicolumn{4}{c}{\textit{Dependent variable: Compliance Percentage Buildings 16:120}} \\ 
\cline{2-5} 
\\[-1.8ex] & \multicolumn{4}{c}{Regression Models} \\ 
\\[-1.8ex] & (1) & (2) & (3) & (4)\\
\hline \\[-1.8ex] 
 DTI &  & 0.003 &  & 0.004 \\ 
  &  & (0.012) &  & (0.013) \\ 
  & & & & \\  
 M1C &  &  & $-$0.015 & $-$0.016 \\ 
  &  &  & (0.067) & (0.068) \\ 
  & & & & \\ 
 Treatment Controls & \checkmark & \checkmark & \checkmark & \checkmark \\ 
  & & & &\\ 
\hline \\[-1.8ex] 
Observations & 49 & 49 & 49 & 49 \\ 
R$^{2}$ & 0.041 & 0.042 & 0.042 & 0.043 \\ 
Adjusted R$^{2}$ & $-$0.023 & $-$0.045 & $-$0.045 & $-$0.068 \\ 
\makecell[l]{Residual Std. \\\hspace{1mm} Error} & \makecell[c]{0.059 \\(df = 45)} & \makecell[c]{0.060 \\(df = 44)} & \makecell[c]{0.060 \\(df = 44)} & \makecell[c]{0.060 \\(df = 43)}\\ 
F Statistic & \makecell[c]{0.634 \\(df = 3; 45)} & \makecell[c]{0.485 \\(df = 4; 44)} & \makecell[c]{0.478 \\(df = 4; 44)} & \makecell[c]{0.390 \\(df = 5; 43)} \\ 
\hline 
\hline \\[-1.8ex] 
\textit{Note:}  & \multicolumn{4}{r}{$^{*}$p$<$0.05; $^{**}$p$<$0.01; $^{***}$p$<$0.001} \\ 
\end{tabular}
\end{table*}

\begin{table}[!htbp] \centering 
  \caption{Study 2 M1C Model Comparison} 
  \label{study2M1CComp} 
\begin{tabular}{@{\extracolsep{5pt}}lcccc} 
\\[-1.8ex]\hline 
\hline \\[-1.8ex] 
Statistic & \multicolumn{1}{c}{(1) Vs. (2)} & \multicolumn{1}{c}{(1) Vs. (3)} & \multicolumn{1}{c}{(2) Vs. (4)} & \multicolumn{1}{c}{(3) Vs. (4)} \\ 
\hline \\[-1.8ex] 
Sum of Sq & $<$0.001 & $<$0.001 &$<$0.001 & $<$0.001 \\ 
F & 0.076 &0.051 & 0.054 & 0.078 \\ 
Pr(\textgreater F) & 0.785 & 0.822 & 0.818 & 0.782\\ 
\hline \\[-1.8ex] 
\end{tabular} 
\end{table}

\begin{table*}[!ht] \centering 
  \caption{Study 3 FC Models} 
  \label{study3FCReg} 
\begin{tabular}{@{\extracolsep{5pt}}lcccc} 
\\[-1.8ex]\hline 
\hline \\[-1.8ex] 
 & \multicolumn{4}{c}{\textit{Dependent variable: Compliance Percentage Buildings 2:45}} \\ 
\cline{2-5} 
\\[-1.8ex] & \multicolumn{4}{c}{Regression Models} \\ 
\\[-1.8ex] & (1) & (2) & (3) & (4)\\
\hline \\[-1.8ex] 
 DTI &  & $-$0.052$^{**}$ &  & $-$0.050$^{**}$ \\ 
  &  & (0.017) &  & (0.017) \\ 
  & & & & \\ 
 FC  &  &  & 0.115$^{***}$ & 0.112$^{***}$ \\ 
  &  &  & (0.031) & (0.030) \\ 
  & & & & \\ 
 Treatment Controls & \checkmark & \checkmark & \checkmark & \checkmark \\ 
  & & & &\\ 
\hline \\[-1.8ex] 
Observations & 148 & 148 & 148 & 148 \\ 
R$^{2}$ & 0.040 & 0.096 & 0.125 & 0.177 \\ 
Adjusted R$^{2}$ & 0.027 & 0.077 & 0.107 & 0.154 \\ 
\makecell[l]{Residual Std. \\\hspace{1mm}Error} &\makecell[c]{0.193 \\(df = 145)} & \makecell[c]{0.188 \\(df = 144)} & \makecell[c]{0.185 \\(df = 144)} & \makecell[c]{0.180 (df = 143)} \\ 
F Statistic & \makecell[c]{3.011 \\(df = 2; 145)} & \makecell[c]{5.078$^{**}$ \\(df = 3; 144)} & \makecell[c]{6.850$^{***}$ \\(df = 3; 144)} & \makecell[c]{7.669$^{***}$ \\(df = 4; 143)} \\ 
\hline 
\hline \\[-1.8ex] 
\textit{Note:}  & \multicolumn{4}{r}{$^{*}$p$<$0.05; $^{**}$p$<$0.01; $^{***}$p$<$0.001} \\ 
\end{tabular} 
\end{table*}

\begin{table}[!htbp] \centering 
  \caption{Study 3 FC Model Comparison} 
  \label{study3FCComp} 
\begin{tabular}{@{\extracolsep{5pt}}lcccc} 
\\[-1.8ex]\hline 
\hline \\[-1.8ex] 
Statistic & \multicolumn{1}{c}{(1) Vs. (2)} & \multicolumn{1}{c}{(1) Vs. (3)} & \multicolumn{1}{c}{(2) Vs. (4)} & \multicolumn{1}{c}{(3) Vs. (4)} \\ 
\hline \\[-1.8ex] 
Sum of Sq & 0.313 & 0.477 & 0.454 & 0.290 \\ 
F & 8.885 & 12.988 & 14.058 & 8.985 \\ 
Pr(\textgreater F) & 0.003 & $<$0.001 & $<$0.001 & 0.003\\ 
\hline \\[-1.8ex] 
\end{tabular} 
\end{table}

\begin{table*}[!ht] \centering 
  \caption{Study 3 AFM Models} 
  \label{study3AFMReg} 
\begin{tabular}{@{\extracolsep{5pt}}lcccc} 
\\[-1.8ex]\hline 
\hline \\[-1.8ex] 
 & \multicolumn{4}{c}{\textit{Dependent variable: Compliance Percentage Buildings 8:45}} \\ 
\cline{2-5} 
\\[-1.8ex] & \multicolumn{4}{c}{Regression Models} \\ 
\\[-1.8ex] & (1) & (2) & (3) & (4)\\
\hline \\[-1.8ex] 
 DTI &  & $-$0.056$^{**}$ &  & $-$0.047$^{**}$ \\ 
  &  & (0.018) &  & (0.016) \\ 
  & & & & \\  
 AFM  &  &  & 0.187$^{***}$ & 0.179$^{***}$ \\ 
  &  &  & (0.032) & (0.032) \\ 
  & & & & \\  
 Treatment Controls & \checkmark & \checkmark & \checkmark & \checkmark \\ 
  & & & &\\ 
\hline \\[-1.8ex] 
Observations & 148 & 148 & 148 & 148 \\ 
R$^{2}$ & 0.037 & 0.098 & 0.220 & 0.263 \\ 
Adjusted R$^{2}$ & 0.024 & 0.080 & 0.204 & 0.242 \\ 
\makecell[l]{Residual Std.\\ \hspace{1mm} Error} & \makecell[c]{0.200 \\(df = 145)} & \makecell[c]{0.194 \\(df = 144)} & \makecell[c]{0.181 \\(df = 144)} & \makecell[c]{0.176 \\(df = 143)} \\ 
F Statistic & \makecell[c]{2.822 \\(df = 2; 145)} & \makecell[c]{5.238$^{**}$ \\(df = 3; 144)} & \makecell[c]{13.577$^{***}$ \\(df = 3; 144)} & \makecell[c]{12.763$^{***}$ \\(df = 4; 143)} \\ 
\hline 
\hline \\[-1.8ex] 
\textit{Note:}  & \multicolumn{4}{r}{$^{*}$p$<$0.05; $^{**}$p$<$0.01; $^{***}$p$<$0.001} \\ 
\end{tabular}
\end{table*}

\begin{table}[!htbp] \centering 
  \caption{Study 3 AFM Model Comparison} 
  \label{study3AFMComp} 
\begin{tabular}{@{\extracolsep{5pt}}lcccc} 
\\[-1.8ex]\hline 
\hline \\[-1.8ex] 
Statistic & \multicolumn{1}{c}{(1) Vs. (2)} & \multicolumn{1}{c}{(1) Vs. (3)} & \multicolumn{1}{c}{(2) Vs. (4)} & \multicolumn{1}{c}{(3) Vs. (4)} \\ 
\hline \\[-1.8ex] 
Sum of Sq & 0.367 & 1.103 & 0.991 & 0.256 \\ 
F & 9.732 & 33.811 & 31.960 & 8.267 \\ 
Pr(\textgreater F) & 0.002 & $<$0.001 & $<$0.001 & 0.005\\ 
\hline \\[-1.8ex] 
\end{tabular} 
\end{table} 

\begin{table*}[!ht] \centering 
  \caption{Study 3 AC-AFM Models} 
  \label{study3ACAFMReg} 
\begin{tabular}{@{\extracolsep{5pt}}lcccc} 
\\[-1.8ex]\hline 
\hline \\[-1.8ex] 
 & \multicolumn{4}{c}{\textit{Dependent variable: Compliance Percentage Buildings 8:45}} \\ 
\cline{2-5} 
\\[-1.8ex] & \multicolumn{4}{c}{Regression Models} \\ 
\\[-1.8ex] & (1) & (2) & (3) & (4)\\
\hline \\[-1.8ex] 
 DTI &  & $-$0.056$^{**}$ &  & $-$0.049$^{**}$ \\ 
  &  & (0.018) &  & (0.016) \\ 
  & & & & \\   
 AC-AFM &  &  & 0.313$^{***}$ & 0.301$^{***}$ \\ 
  &  &  & (0.053) & (0.052) \\ 
  & & & & \\   
 Treatment Controls & \checkmark & \checkmark & \checkmark & \checkmark \\ 
  & & & &\\ 
\hline \\[-1.8ex] 
Observations & 148 & 148 & 148 & 148 \\ 
R$^{2}$ & 0.037 & 0.098 & 0.223 & 0.269 \\ 
Adjusted R$^{2}$ & 0.024 & 0.080 & 0.207 & 0.249 \\ 
\makecell[l]{Residual Std. \\\hspace{1mm} Error} & \makecell[c]{0.200 \\(df = 145)} & \makecell[c]{0.194 \\(df = 144)} & \makecell[c]{0.180 \\(df = 144)} & \makecell[c]{0.175 \\(df = 143)} \\ 
F Statistic & \makecell[c]{2.822 \\(df = 2; 145)} & \makecell[c]{5.238$^{**}$ \\(df = 3; 144)} & \makecell[c]{13.767$^{***}$ \\(df = 3; 144)} & \makecell[c]{13.164$^{***}$\\(df = 4; 143)} \\ 
\hline 
\hline \\[-1.8ex] 
\textit{Note:}  & \multicolumn{4}{r}{$^{*}$p$<$0.05; $^{**}$p$<$0.01; $^{***}$p$<$0.001} \\ 
\end{tabular}
\end{table*}

\begin{table}[!htbp] \centering 
  \caption{Study 3 AC-AFM Model Comparison} 
  \label{study3ACAFMComp} 
\begin{tabular}{@{\extracolsep{5pt}}lcccc} 
\\[-1.8ex]\hline 
\hline \\[-1.8ex] 
Statistic & \multicolumn{1}{c}{(1) Vs. (2)} & \multicolumn{1}{c}{(1) Vs. (3)} & \multicolumn{1}{c}{(2) Vs. (4)} & \multicolumn{1}{c}{(3) Vs. (4)} \\ 
\hline \\[-1.8ex] 
Sum of Sq & 0.367 & 1.116 & 1.028 & 0.278 \\ 
F & 9.732 & 34.359 & 33.403 & 9.046 \\ 
Pr(\textgreater F) & 0.002 & $<$0.001 & $<$0.001 & 0.003\\ 
\hline \\[-1.8ex] 
\end{tabular} 
\end{table}

\begin{table}[!htbp] \centering 
  \caption{DTI as a Predictor of First Compliance Choice} 
  \label{DTIhypTab} 
\begin{tabular}{@{\extracolsep{5pt}}lccc} 
\\[-1.8ex]\hline 
\hline \\[-1.8ex] 
 & \multicolumn{3}{c}{\textit{Dependent variable: First Choice}} \\ 
\cline{2-4} 
\\[-1.8ex] & \multicolumn{3}{c}{Regression Models} \\ 
\\[-1.8ex] & Study 1 & Study 2 & Study3\\ 
\hline \\[-1.8ex] 
 DTI & 0.024 & 0.002 & $-$0.017 \\ 
  & (0.021) & (0.092) & (0.047) \\ 
  & & &  \\ 
 Controls & \checkmark & \checkmark & \checkmark \\ 
  & & & \\ 
\hline \\[-1.8ex] 
Observations & 198 & 48 & 148 \\ 
R$^{2}$ & 0.062 & 0.145 & 0.003 \\ 
Adjusted R$^{2}$ & 0.022 & 0.065 & $-$0.018 \\ 
Residual Std. Error & 0.415 & 0.434 & 0.499 \\ 
  & (df = 189) & (df = 43) & (df = 144) \\ 
F Statistic & 1.562 & 1.820 & 0.141 \\ 
  & (df = 8; 189) & (df = 4; 43) & (df = 3; 144) \\ 
\hline 
\hline \\[-1.8ex] 
\textit{Note:}  & \multicolumn{3}{r}{$^{*}$p$<$0.1; $^{**}$p$<$0.05; $^{***}$p$<$0.01} \\ 
\end{tabular} 
\end{table} 

\end{subappendices}
\end{document}